\documentclass[11pt]{article} 
\usepackage[utf8]{inputenc}
\usepackage{geometry}
\usepackage{xcolor}
\usepackage{authblk}
\usepackage{amsmath}
\usepackage{graphicx}
\usepackage{caption}
\usepackage{comment}
\usepackage{accents} 
\usepackage{float}
\usepackage{lineno}
\usepackage{setspace}
\usepackage[authoryear]{natbib}
\usepackage[colorlinks=true, allcolors=blue]{hyperref}
\usepackage{cleveref}
\newcommand{\ie}{i.e., }

\title{Using theory-driven Integrated Population Models to evaluate competitive outcomes in stage-structured systems}
\author[1,2]{Matthieu Paquet}
\author[1]{Frédéric Barraquand}
\affil[1]{\normalsize Institute of Mathematics of Bordeaux\\

University of Bordeaux, CNRS, Bordeaux INP, Talence, France}

\affil[2]{\normalsize Theoretical and Experimental Ecology Station (SETE), CNRS, Moulis, France}
\date{\vspace{-5ex}}

\begin{document}

\maketitle


\thispagestyle{empty}

\begin{abstract}
Predicting competitive outcomes typically requires fitting dynamical models to data, from which interaction strengths and coexistence indicators such as invasion criteria can be produced. Methods that allow to propagate parameter uncertainty are particularly indicated. These should ideally allow for competition between and within species at various life-stages, and make the best out of multiple data sources, each of which can be relatively scarce by statistical standards. Here, we embed a mathematical model of stage-structured competition between two species, producing analytical invasion criteria, into a two-species Integrated Population Model. The community-level IPM allows to combine counts, capture-recapture, and fecundity data into a single statistical framework, and the Bayesian formulation of the IPM fully propagates parameter uncertainty into invasion criteria. Model fitting demonstrates that we can correctly predict coexistence through reciprocal invasion when present, but that interaction strengths are not always estimable, depending on the prior chosen. Our competitive exclusion scenario is shown to be harder to identify, although our model allows to at least flag this scenario as uncertain rather than mistakenly present it as coexistence. Our results confirm the importance of accounting for uncertainty in the prediction of competitive outcomes. 
\end{abstract}


Correspondence to \url{matthieu.paquet@sete.cnrs.fr}
\newpage

\section{Introduction}
Two-species coexistence is typically predicted by Lotka--Volterra models when intraspecific competition is stronger than interspecific competition \citep{chesson2000mechanisms,letten2017linking}. 
However, any deviation from the unstructured two-species Lotka--Volterra framework, such as adding more species \citep{barabas2016effect} or more ages / stages \citep{kohyama1992size,moll2008competition,fujiwara2011coexistence} can make the mapping between species interaction strengths and competitive outcomes (coexistence, exclusion, priority effects) much less straightforward. In stage- or age-structured populations, inferring coexistence therefore requires further investigations by calculating invasion criteria, or running simulations \citep{moll2008competition,fujiwara2011coexistence}. Yet, it is of particular interest to investigate coexistence in such stage-structured systems, as in most communities some age- or stage-structure will be visible and potentially mediate the susceptibility of individuals to competition, be those communities made of trees, birds, fishes or insects.  Theoretical models have actually shown that the presence of age structure itself could contribute to species coexistence \citep{moll2008competition,fujiwara2011coexistence}, at least in systems with few, similar, strongly competing species \citep{bardon2023effects}. While these advances in modelling competing stage structured populations are shedding light on our understanding of conditions promoting species coexistence, whether and under which circumstances these models' parameters could be estimated using real data remains unknown. As dynamic population models with interactions are parameter-rich, there are expected challenges associated with fitting such models to data and estimate species interactions and coexistence \citep{paniw2023pathways}. One of them being that some parameters such as species interaction coefficients can be hard to estimate from population count data alone \citep{barraquand2019integrating}.

Combining several datasets into integrated models can help estimating parameters that would not be identifiable if datasets were analysed separately \citep{cole2016parameter}, including for estimating species interactions \citep{barraquand2019integrating}. Stage-structured multispecies integrated population models (IPMs) have been developed to estimate species interactions \citep{peron2012integrated,barraquand2019integrating,queroue2021multispecies,paquet2022assessing,viollat2024bottom}. However, simulation-based evaluations have been performed in a trophic interaction set-up \citep{barraquand2019integrating,paquet2022assessing} and whether these models provide reliable estimates of competitive interactions is still poorly known, particularly in cases where both intra- and interspecific interactions act on the same vital rates. Moreover, they rely on phenomenological functions (using log and logit links), which is very convenient from a statistical perspective but has some drawbacks for ecological interpretation. Indeed, stage-structured models with log and logit links in vital rates \citep{moll2008competition,peron2012integrated} do not allow for immediate assessment of competitive outcomes from interaction strengths, and analytically tractable invasion criteria are not available either, which means that coexistence can only be assessed by simulating the fitted models. More theory-driven competitive functional forms, and resulting population models, could be used to infer populations' coexistence. This should be particularly interesting in models for which
invasion criteria have been obtained analytically \citep{fujiwara2011coexistence,bardon2023effects}. 
Finally, when modelled in a Bayesian framework, such two-species theory-driven IPMs could allow to estimate invasion criteria (hence species coexistence or exclusion) together with their uncertainty, potentially even when all model parameters are not identifiable individually \citep[by analogy to epidemiological models,][]{kao2018practical}. This need to correctly propagate uncertainty has been recently reflected in the plant coexistence literature \citep{bowler2022accounting,armitage2024remain}. We therefore draw here on the development of such stochastic multispecies IPMs in animal ecology to improve the evaluation of competitive outcomes, such as coexistence or exclusion in plant and animal ecology alike. 

In the following, we propose a two-species Integrated Population Model---a dynamic, demographically structured model relying on multiple data sources---to estimate competitive interaction coefficients and evaluate coexistence or exclusion of two (potentially) competing stage-structured populations. In this model, adults can affect both adults and juveniles of either species, so that both fecundities and juvenile survival rates can depend upon adult densities of both species. Unlike previous multispecies IPMs relying on Ricker-based functional forms (log link functions), this model encapsulates a Beverton-Holt two-species juvenile-adult population model, for which analytical expressions of invasion criteria have been obtained \citep{bardon2023effects}. Using four sets of parameters (\ie three contrasting sets leading to coexistence through different pathways, and one leading to one species excluding the other), we assess the identifiability and accuracy of model parameters estimates, including species interactions, as well as our ability to estimate the mathematically-derived invasion criteria.

\section{Methods}

\subsection{Description of the two-species two-stages stochastic model}

The model is based on \citet{bardon2023effects} and their stage-structured difference equation is then embedded in a stochastic framework to account for demographic stochasticity (similarly to \citealt{barraquand2019integrating}). For each species we used a post-breeding census model where we have the expected number of breeders at each time step:
\begin{equation}\label{eq:adultaddition}
n_a(t+1) = n_a^{\text{old}}(t+1) + n_a^{\text{new}}(t+1)
\end{equation}
where $n_a^{\text{old}}(t+1)$ represents the number of adults (breeders) that were already adults at the previous time step, and $n_a^{\text{new}}(t+1)$ represent the number adults that were juveniles at the previous time step. We model demographic stochasticity using Binomial and Poisson distributions, so that
\begin{equation}\label{eq:oldadultstoch}
n_a^{\text{old}}(t+1) \sim \text{Binomial}\left(s_a,n_a(t)\right) ,
\end{equation}
with $s_a$ the adult yearly survival, and
\begin{equation}\label{eq:newadultstoch}
n_a^{\text{new}}(t+1) \sim \text{Binomial}\left(s_j(t),n_j(t)\right) 
\end{equation}
with $s_j(t)$ the juvenile yearly survival (that is time dependent, see below), and $n_j(t)$ the number of juveniles at time t. Finally for the number of juveniles we have
\begin{equation}\label{eq:Juvenilestoch}
n_j(t+1) \sim \text{Poisson}\left(f(t)n_a(t)\right),
\end{equation}
with $f(t)$ representing fecundity (the number of offspring of the modelled sex at each time step). The population model used here is actually a slight simplification of the model presented in \citet{bardon2023effects}, as the maturation rate ($\gamma$) is set to 1, that is, all surviving juveniles become adults (none stay in the juvenile stage).

Regarding the density-dependent parameters, for species 1 we have:
\begin{equation}\label{eq:survlink}
s_{1j}(t) = \frac{\phi_1}{1+\beta_{11} n_{1a}(t) + \beta_{12} n_{2a}(t)}
\end{equation}
where $\phi_1$ represents the maximum juvenile survival rate, that is survival when (hypothetically) the number of adults of both species $n_{1a}(t)$ and $n_{2a}(t)$ is zero. Parameters $\beta_{11}$ and $\beta_{12}$ represent the negative effect of $n_{1a}$ and $n_{2a}$, that is, the strength of intra- and inter-specific density-dependence respectively. Similarly, for the fecundity of species 1 we have:
\begin{equation}\label{eq:feclink}
  f_1(t) = \frac{\pi_1}{1+\alpha_{11} n_{1a}(t) + \alpha_{12} n_{2a}(t)}  .\end{equation}
where $\pi_1$ represents the maximum fecundity when $n_{1a}(t)$ and $n_{2a}(t)$ are zero, and $\alpha_{11}$ and $\alpha_{12}$ represent the negative effects of $n_{1a}$ and $n_{2a}$, respectively. Because we are in a post-breeding census, formally $\pi_1$ should be divided by $s_a$ to be interpreted as the true maximum fecundity; we kept the notations from previous theoretical papers for simplicity and consistency. \\
  
Then by symmetry, for species 2 we have:

\begin{equation}\label{eq:survlink2}
s_{2j}(t) = \frac{\phi_2}{1+ \beta_{22} n_{2a}(t) + \beta_{21} n_{1a}(t)}
\end{equation}
and
\begin{equation}\label{eq:feclink2}
  f_2(t) = \frac{\pi_2}{1+\alpha_{22} n_{2a}(t) + \alpha_{21} n_{1a}(t)}  .\end{equation}
  
\subsubsection{Count data}

Here we assume that juvenile and adult individuals are distinguishable throughout the population census (i.e., what produces the count data), which is particularly likely for the post-breeding census model considered here, since the young are counted right after birth and therefore are very unlikely to resemble already their parents. We further assume that the observation error around the number of juveniles and adults follows a Poisson distribution, simulating a study case where we sample a given area with a given underlying density of individuals. For juvenile counts of species 1 we have:
\begin{equation}\label{eq:countyoung1}
y_{1j}(t)\sim \text{Poisson}\left(n_{1j}(t) \right)
\end{equation}
and for adults:
\begin{equation}\label{eq:countadult1}
y_{1a}(t)\sim \text{Poisson}\left(n_{1a}(t) \right)
.\end{equation}

\subsubsection{Survival data}

We simulated and fitted the capture-mark-recapture data in the m-array format, using a multinomial likelihood \citep{burnham1987design}. The data is in the form of two $(T-1) \times T$ matrices $\mathbf{M}^{J}$ and $\mathbf{M}^{A}$, one for each age class, with $\mathbf{M}^{(a)}=(m^{(a)}_{t,l})$, with $m^{(a)}_{t,l}=0, \; \forall l<t$, where $T$ is the total number of years of capture recapture history.
$m^{(a)}_{t,t}$ is the number of individuals captured and released at age class $(a)$ at time $t$ that were recaptured the following year, and the last column $m^{(a)}_{t,T}$ is the number of individuals captured at age class $(a)$ at time $t$ that were never recaptured. We then have:
\begin{equation}
\mathbf{m}^{(a)}_{t,\bullet}=(m^{(a)}_{t,t},m^{(a)}_{t,t+1},\ldots,m^{(a)}_{t,T})\sim \text{Multinomial}\left( R^{(a)}_t,({\theta}^{(a)}_{t,t},\ldots,{\theta}^{(a)}_{t,T}) \right)
\end{equation} with $R^{(a)}_t= \sum_{k=t}^{T}m^{(a)}_{t,k}$ the number of individuals of age class $(a)$ captured at time $t$.

For juveniles, diagonal elements of the $\boldsymbol{\theta}^J$ matrix write:
\[ \theta^J_{t,t}=s_j(t)p,\]
with $s_j(t)$ the first year (i.e. juvenile) survival probability from year $t$ to year $t+1$ (for the species considered), and $p$ the recapture probability set as constant among years and age classes, and for $t<l<T$
\[ \theta^J_{t,l}=s_j(t)s_a^{l-t} (1-p)^{l-t}p,\]

with $s_a$ the adult survival probability (constant across years) for the species considered. The last element pertains to individuals never recaptured
\[\theta^J_{t,T}=1-\sum_{k=t}^{T-1}\theta^J_{t,k}.\]

Similarly, for $\boldsymbol{\theta}^A$, the above mentioned equations are identical to the exception that $s_j$ is replaced by $s_a$, which leads to:
\[ \theta^A_{t,t}=s_ap\] for the diagonal elements of the $\boldsymbol{\theta}^A$ matrix,
and for $t<l<T$:
\[ \theta^A_{t,l}=s_a^{l-t+1} (1-p)^{l-t}p.\]
The last element again pertains to individuals never recaptured
\[\theta^A_{t,T}=1-\sum_{k=t}^{T-1}\theta^A_{t,k}.\]

\subsubsection{Fecundity data}

Fecundity was modelled using a Poisson regression:
\begin{equation}
F_t\sim \text{Poisson}(n_{\text{rep},t}2f(t))
\end{equation} with $F_t$ the total number of offspring counted, $n_{\text{rep},t}$ the number of surveyed broods/litters per year, and $f(t)$ the expected number of offspring females per adult female each year $t$ (assuming a sex ratio of 1/1).

\subsection{Description of the four sets of parameter values}
We used the three sets of parameters leading to species coexistence (if ignoring demographic stochasticity) that were explored in \citet{bardon2023effects}. Parameter set 1 leads to a classic coexistence which is reflected by both $\alpha$ and $\beta$ coefficients (similar to the intra $>$ interspecific competition rule of unstructured Lotka--Volterra models). Parameter set 2 is by contrast an \textit{emergent} coexistence state: competition coefficients $\alpha$ alone would lead to species 1 winning, competition coefficients $\beta$ would lead to species 2 winning, but having these non-zero $\alpha$ and $\beta$ values in the same model leads to 1 and 2 coexisting.  Parameter set 3 represents classical coexistence through $\alpha$ and a priority effect through $\beta$, leading to emergent coexistence at the full-model level. 

In addition, we used one set of parameters leading to no coexistence (extinction of species 2) in order to assess the performance of the model in estimating the coexistence criteria in such case (\Cref{table:paramvalues_dd}).

\begin{table}[H]
\begin{center}
\caption{Density-dependence parameters with their values for each of the four parameter sets.
}\label{table:paramvalues_dd}
\begin{tabular}{|c|llll|}
\hline
Parameter & Set 1 & Set 2 & Set 3 & Set 4\\
\hline
$\alpha_{11}$ &$0.1$&$0.1$&$0.1$&$0.1$ \\
$\alpha_{12}$ &$0.05$&$0.02$&$0.043$&$0.1$ \\
$\alpha_{21}$ &$0.06$&$0.112$&$0.035$&$0.1$ \\
$\alpha_{22}$ &$0.1$&$0.1$&$0.1$&$0.1$ \\
$\beta_{11}$ &$0.1$&$0.1$&$0.1$&$0.1$ \\
$\beta_{12}$ &$0.06$&$0.125$&$0.155$&$0.1$ \\
$\beta_{21}$ &$0.06$&$0.01$&$0.165$&$0.1$ \\
$\beta_{22}$ &$0.1$&$0.1$&$0.1$&$0.1$ \\
\hline
\end{tabular}
\end{center}
\end{table}

Other parameter values were identical for the four parameter sets (\Cref{table:paramvalues_constant}, recapture probabilities were $p_1=p_2=0.7$ and initial stage specific abundances $n_{1j}(1)=n_{1a}(1)=n_{2j}(1)=n_{2a}(1)=100$).
\begin{table}[H]
\begin{center}
\caption{Other demographic parameters with their values (common to all four parameter sets).
}\label{table:paramvalues_constant}
\begin{tabular}{|c|c|}
\hline
Parameter & Value \\
\hline
$s_{1a}$&$0.5$ \\
$s_{2a}$&$0.6$ \\
$\phi_1$&$0.5$\\
$\phi_2$&$0.4$\\
$\pi_1$&$30$\\
$\pi_2$&$25$\\
\hline
\end{tabular}
\end{center}
\end{table}

\subsection{Sample sizes of the simulated datasets}

\begin{figure}[H]
\includegraphics[width=0.95\linewidth]{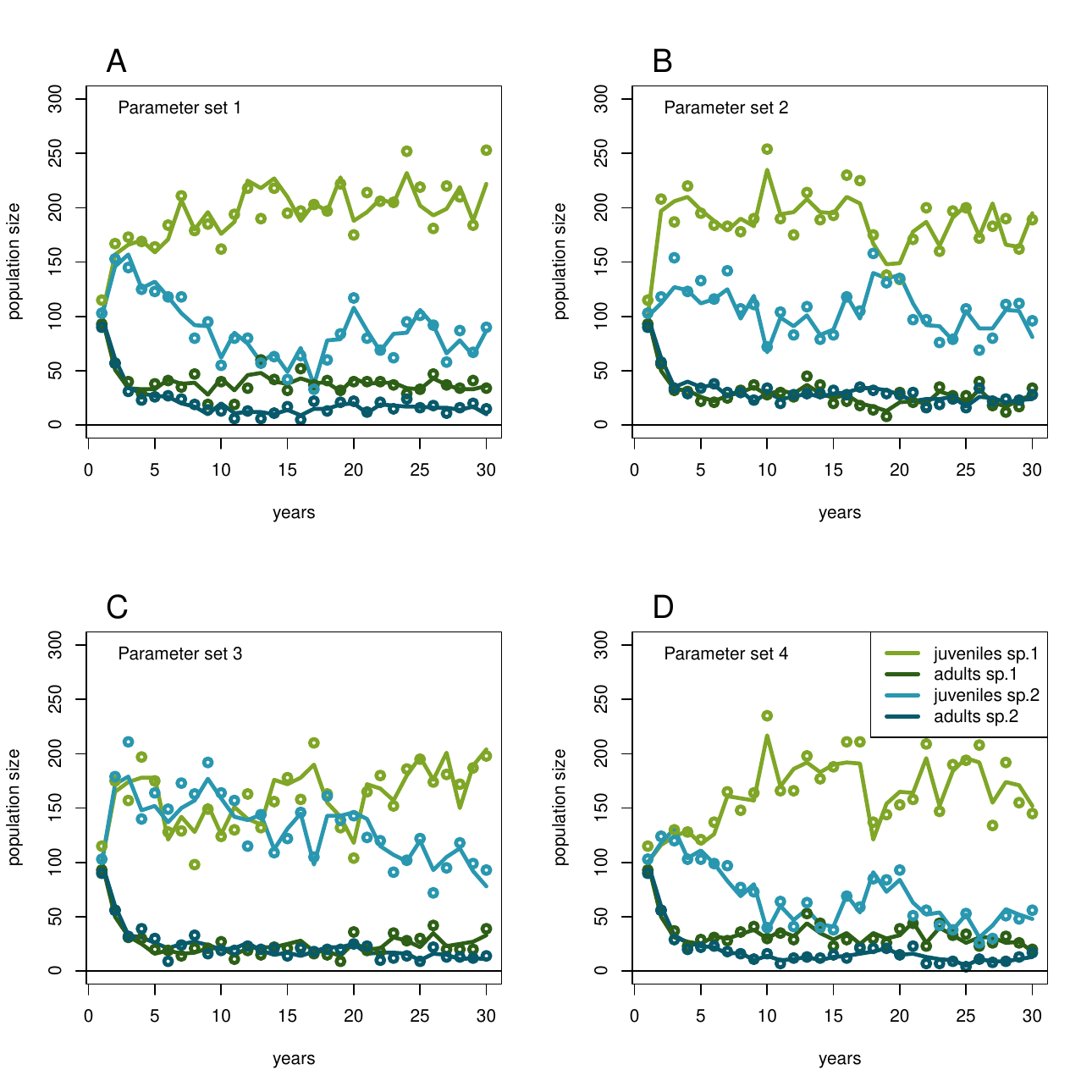}
\centering
\caption{Examples (first simulations of each series of 100) of time series of age specific abundances and counts for each parameter set. Light green lines and circles represent $n_{1j}$ and $y_{1j}$ respectively, dark green lines and circles represent $n_{1a}$ and $y_{1a}$, light blue lines and circles represent $n_{2j}$ and $y_{2j}$, and dark blue lines and circles represent $n_{2a}$ and $y_{2a}$.}
\label{fig:timeseries}
\end{figure}

For both species and all four simulation scenarios (including 100 simulations for each of the four parameter sets), we used a study period of $T=30$ years, a yearly number of monitored broods/litters $n_{\text{rep},t}=50$, and a yearly number of marked juveniles $R^{J}_t=100$. We chose these sample sizes based on a previous 2-species IPM which provided accurate parameter estimates with satisfactory precision \citep{paquet2022assessing}. These represent a substantial yet attainable survey effort in well-studied taxa such as small birds. Four examples (one per parameter set) of simulated age specific population sizes and their associated observed count data are shown in \Cref{fig:timeseries}.

\subsection{Prior specification}

\begin{figure}[H]
\includegraphics[width=0.95\linewidth]{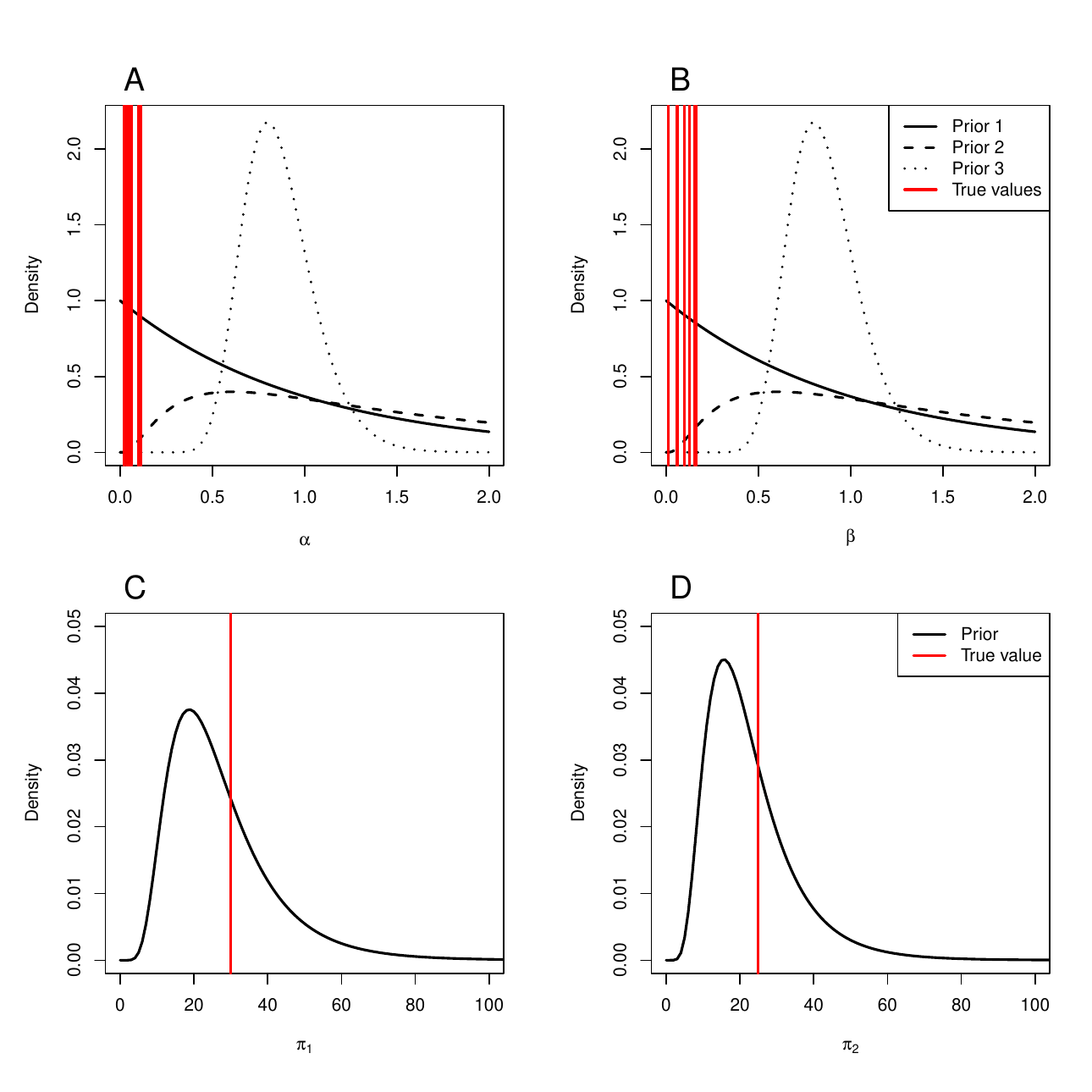}
\centering
\caption{Density of the three priors used for the density-dependence coefficients $\alpha_{i,j}$ (panel A) and $\beta_{i,j}$ (panel B). Prior 1 (full line) = $\exp(1)$, prior 2 (dashed line) = $\mathcal{L N}(0.5,1)$, and prior 3 (dotted line) = $\mathcal{LN}(\ln(0.8)+0.05,0.05)$. Red lines show all true values used for $\alpha_{i,j}$ and $\beta_{i,j}$ across the four parameter sets (see \Cref{table:paramvalues_dd} for their values). Panel C and D: Prior density (black lines) and true values (red lines) for maximum expected fecundity parameters $\pi_i$ of species 1 (C) and species 2 (D).}
\label{fig:priors}
\end{figure}

To assess the sensitivity of the estimation of interaction parameters ($\alpha_{i,j}$ and $\beta_{i,j}$) and invasion criteria to the choice of prior probability densities, we fitted three models to each dataset, each using a different type of prior for $\alpha_{i,j}$ and $\beta_{i,j}$ (\Cref{fig:priors}). First we used exponential priors $\theta \sim \text{Exp}(1)$, hereafter prior 1, which is a relatively vague prior with a mode (\ie most probable value) at zero. Second we used a log-normal prior $\theta \sim \mathcal{L N}(0.5,\sigma^2=1)$, hereafter prior 2, which is also a vague prior with a mode around $0.6$ ($\text{Exp}(-0.5)$). Finally, we used $\theta \sim \mathcal{LN}(\log(0.8)+\sigma^2,\sigma^2) = \mathcal{LN}(\log(0.8)+0.05,0.05)$, hereafter prior 3, with mode $0.8$, which is a rather informative prior where most of the probability distribution is concentrated away from the true values (\Cref{fig:priors}). While priors 1 and 2 were chosen because they both appeared to us as reasonable choices, prior 3 was specifically chosen as an extreme case for our prior sensitivity diagnostic.

For setting up priors of maximum expected fecundity parameters $\pi_i$, preliminary analyses using uniform priors led to issues with posterior samples systematically accumulating at the higher boundaries defined. To avoid this issue and having to define an arbitrary higher boundary we used log-normal distributions (\Cref{fig:priors}):
\begin{equation}
 \pi_i^{prior} \sim \mathcal{L N} (\log(m_{\pi_i^{\text{prior}}})-\sigma_{\text{prior}}^2/2, \sigma_{\text{prior}}^2=0.25)
\end{equation}
with \begin{equation} \label{eq:priorpi}
    m_{\pi_i^{\text{prior}}}=\frac{\pi_i}{1+\alpha_{i,i}},
\end{equation}
that is, the maximum expected fecundity value observable $f_i$ without adult competitors of the other species $j$ ($n_{j\neq i,a}=0$) and with one adult of the modelled sex ($n_{i,a}=1$). This is supposed to represent a case where one would make an informed choice and use the highest known fecundity value of a population/species as a prior mean value for max fecundity. We chose $\sigma_{prior}=0.5$ as it seems to provide a density of reasonable width with true values falling well within the prior mass (\Cref{fig:priors}), but note that this prior may be informative if uncertainty around the estimation of $\pi$ is high.

Other prior probabilities, for maximum expected survival $\phi_i$ and recapture $p$, were drawn from uniform distributions $\text{Unif}(0,1)$. For the initial stage-specific population sizes of both species, we used $\mathcal{N}(100,\sigma^2=100)$ priors  (rounded and truncated to be positive).

\subsection{Model fitting}

Data were both simulated and fitted using the Nimble R package \citep[][version 0.13.1]{rcitation,nimble-article:2017,nimble-software:2023}.
For each simulated dataset, we fitted the same multispecies IPM that was used to generate the data.

Two MCMC chains were run for 15100 iterations and we sampled the last 15000 iterations every 10\textsuperscript{th} iteration leading to 3000 posterior samples saved per dataset.
We used 2 sets of initial values for parameters, one for each chain (same for all first chains and same for all second chains). For maximum fecundity $\pi_1$ and $\pi_2$ we used $m_{\pi_i^{\text{prior}}}$ from \Cref{eq:priorpi} for both chains. This is to reflect what would be the maximum value one could observe (minus stochasticity due to the Poisson process). So here we use a value smaller than the true value for both species (\ie $30/1.1$ and $25/1.1$ for species 1 and species 2 respectively). For initial age specific population sizes we used $\mathcal{N}(100,\sigma=10)$ to simulate the initial values. For the other parameters we used uniform distributions with realistic and rather restrained ranges.

We assessed convergence and mixing of the chains by calculating the potential scale reduction factor ($\hat{R}$, \citealt{brooks1998general,gelman1992inference}) and effective sample size ($n_{\text{eff.}}$)using the \texttt{gelman.diag()} and the \texttt{effectiveSize()} functions of the \texttt{coda} package \citep[][version 0.19-4]{codaplummer}. We only used outputs from models for which all $\alpha_i$, $\beta_i$, $\phi_i$, and $\pi_i$ had $\hat{R}<1.1$ and $n_{\text{eff.}}>50$, that is, $1195/1200$ model runs (maximum 1 run discarded per Parameter set $\times$ Prior combination).

To improve their mixing and minimize their posterior correlations, parameters within the same density-dependent functions were block sampled [$(\pi_1,\alpha_{11},\alpha_{12})$, $(\pi_2,\alpha_{22},\alpha_{21})$, $(\phi_1,\beta_{11},\beta_{12})$ and $(\phi_2,\beta_{22},\beta_{21})$] using automated factor slice samplers \citep{tibbits2014automated,ponisio2020one}.

The computer code is provided at
\url{https://github.com/MatthieuPaquet/IPM_competition_2_species_age_structured}.

\subsection{Computation of invasion criteria}

To estimate whether the two species are expected to i) coexist, ii) exclude each other, or iii) exhibit priority effects, we computed mathematically-derived invasion criteria from the posterior samples of the model parameters (enabling us to propagate uncertainty around the estimates). The detailed derivation of the invasion criteria (denoted $\mathcal{I}_1$ and $\mathcal{I}_2$ here) are provided in \citet{bardon2023effects}, where they found that the condition for stability when species 1 is absent and species 2 is resident is: 

\begin{equation}
    \mathcal{I}_1 = \frac{C_1}{1 + \beta_{12} n^*_{2a}} + \frac{D_1}{(1 + \beta_{12} n^*_{2a}) (1 + \alpha_{12} n^*_{2a})}<1
\end{equation}
with
\begin{equation}
    D_1 = \frac{\pi_1 \gamma_1 \phi_1}{1 - s_{1a}}
,\end{equation}

\begin{equation}
    C_1 = (1-\gamma_1)\phi_1
\end{equation}
and
\begin{equation}
    n^*_{2a} = \frac{(\alpha_{22} C_2 - \alpha_{22} - \beta_{22}) + \sqrt{(- \alpha_{22} C_2 + \alpha_{22} + \beta_{22})^2 -   4 \alpha_{22} \beta_{22} (1-C_2-D_2)}}{2 \alpha_{22} \beta_{22}}
\end{equation}
with
\begin{equation}
    D_2 = \frac{\pi_2 \gamma_2 \phi_2}{1 - s_{2a}},
\end{equation}
\begin{equation}
    C_2 = (1-\gamma_2)\phi_2,
\end{equation}
and
\begin{equation}
    C_2 + D_2 <1.
\end{equation}

Since the stage transition probabilities $\gamma_1$ and $\gamma_2$ were fixed to 1 in our model (and consequently $C_1=0$ and $C_2=0$), these formula simplify and the invasion criterion for species 1 is:
\begin{equation}
    \mathcal{I}_1 = \frac{D_1}{(1 + \beta_{12} n^*_{2a}) (1 + \alpha_{12} n^*_{2a})}
\end{equation}
with
\begin{equation}
    D_1 = \frac{\pi_1 \phi_1}{1 - s_{1a}}
\end{equation}
and
\begin{equation}
    n^*_{2a} = \frac{- \alpha_{22} - \beta_{22} + \sqrt{(\alpha_{22} + \beta_{22})^2 -   4 \alpha_{22} \beta_{22} (1-D_2)}}{2 \alpha_{22} \beta_{22}}
\end{equation}
with
\begin{equation}
    D_2 = \frac{\pi_2 \phi_2}{1 - s_{2a}}>1.
\end{equation}
If both $\mathcal{I}_1>1$ and $\mathcal{I}_2>1$, then species are expected to coexist. If $\mathcal{I}_1>1$ and $\mathcal{I}_2<1$ then species 1 is expected to exclude species 2 (and \emph{vice versa}), and if both $\mathcal{I}_1<1$ and $\mathcal{I}_2<1$ a priority effect is expected.

\section{Results}

\subsection{Estimating intra- and interspecies density-dependence parameters}

\begin{figure}[H]
\includegraphics[width=0.95\linewidth]{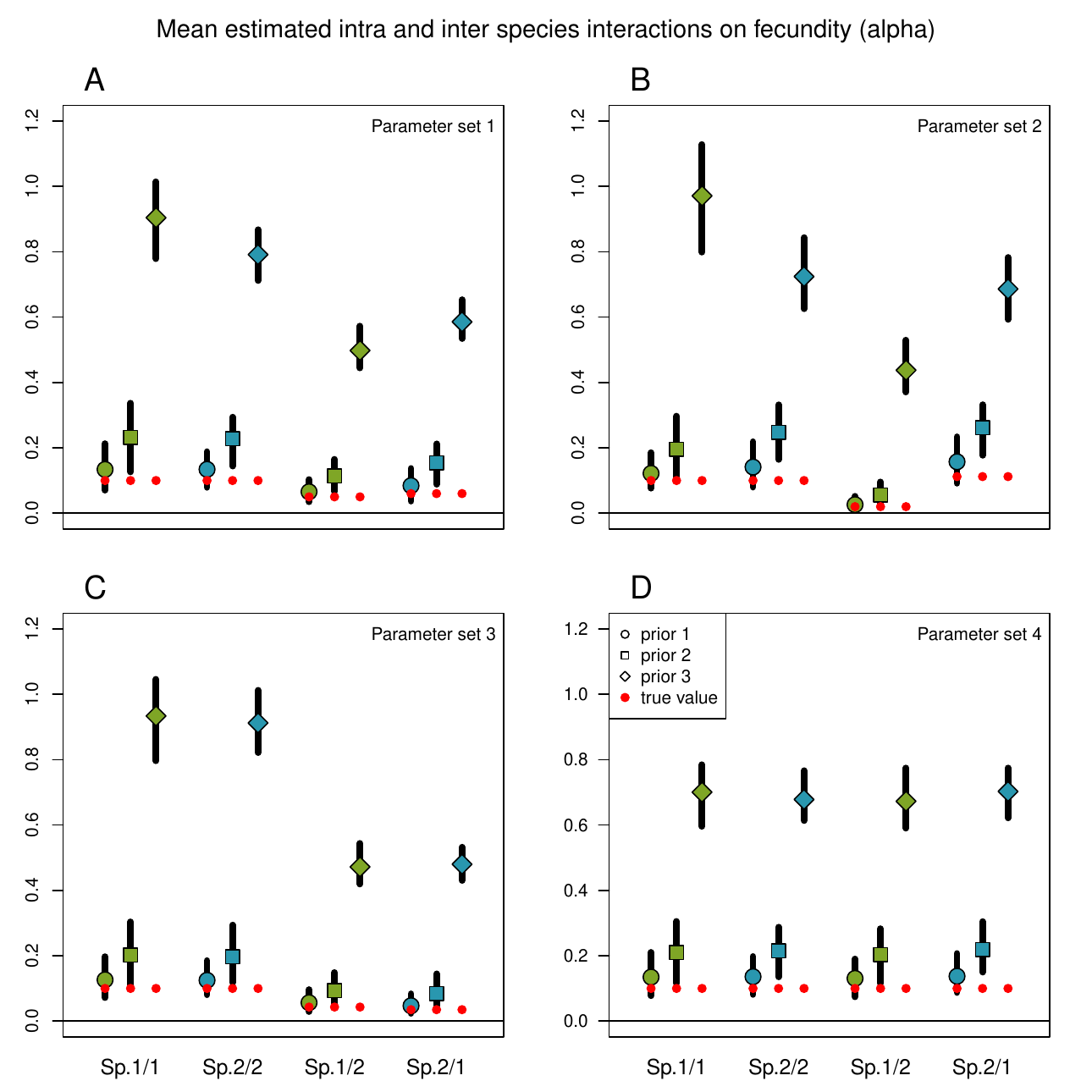}
\centering
\caption{Point estimates (\ie posterior means) of intra- and inter-species interactions on fecundity ($\alpha_{i,j}$) estimated under each parameter set (panel A = parameter set 1; panel B = parameter set 2; panel C = parameter set 3; panel D = parameter set 4). Vertical lines represent the 95\% intervals of the (typically 100) posterior means, and open dots represent their means (filled in green for effects of species 1 and in blue for effects of species 2). Round dots represent estimates when prior 1 was applied, square dots when prior 2 was applied, and diamond dots when prior 3 was applied to all $\alpha_{i,j}$ and $\beta_{i,j}$. Red dots represent the true values. ``Sp.1/1'' refers to $\alpha_{1,1}$, that is, the negative effect of the number of adults of species 1 on its own fecundity, ``Sp.1/2'' refers to the negative effect of the number of adults of species 1 on the fecundity of species 2, and so on.}
\label{fig:alphaest}
\end{figure}

\begin{figure}[H]
\includegraphics[width=0.95\linewidth]{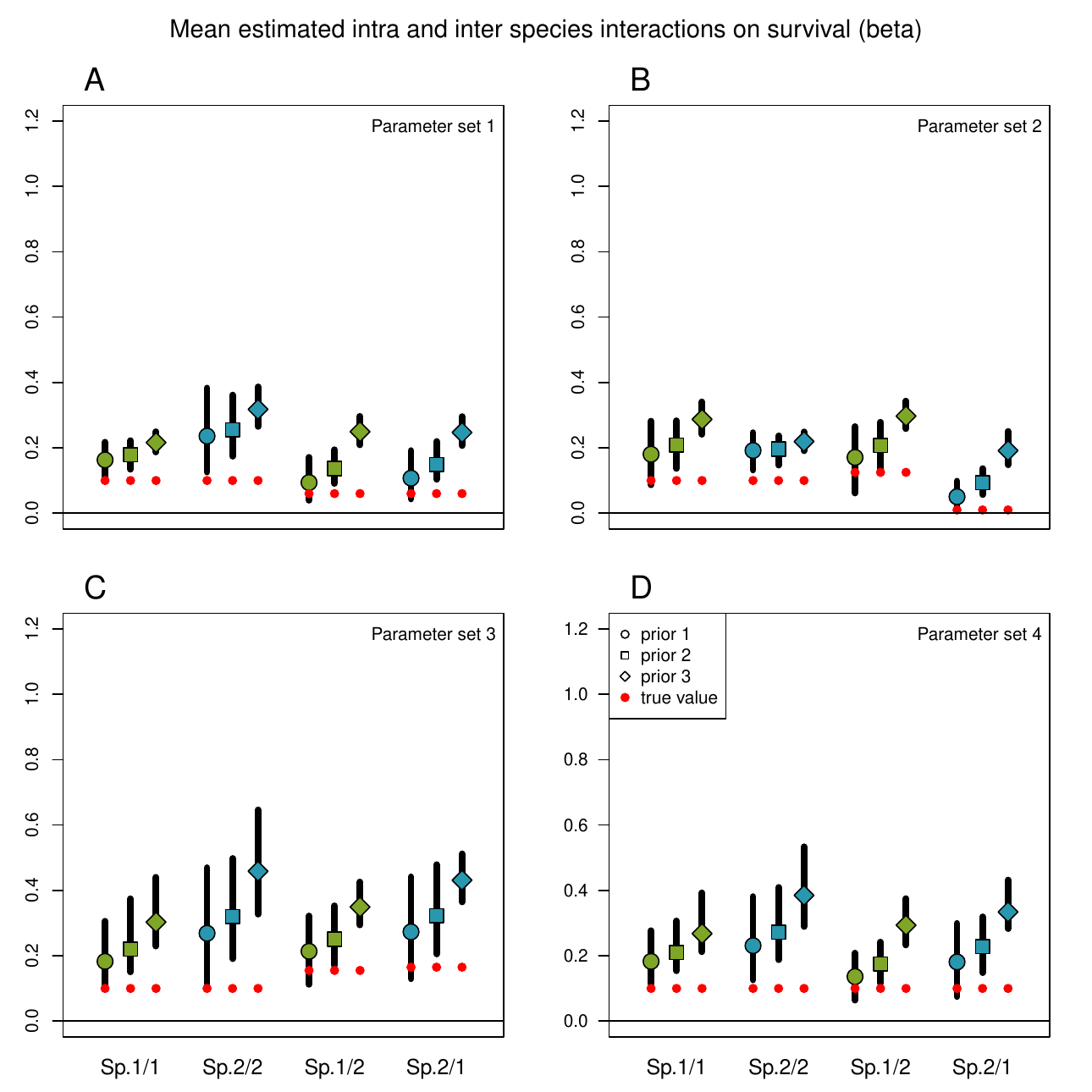}
\centering
\caption{Point estimates (\ie posterior means) of intra- and inter-species interactions on survival ($\beta_{i,j}$) estimated under each parameter set (panel A = parameter set 1; panel B = parameter set 2; panel C = parameter set 3; panel D = parameter set 4). Vertical lines represent the 95\% intervals of the (typically 100) posterior means, and open dots represent their means (filled in green for effects of species 1 and in blue for effects of species 2). Round dots represent estimates when prior 1 was applied, square dots when prior 2 was applied, and diamond dots when prior 3 was applied to all $\alpha_{i,j}$ and $\beta_{i,j}$. Red circles represent the true values. ``Sp.1/1'' refers to $\beta_{1,1}$, that is, the negative effect of the number of adults of species 1 on the survival of juveniles of the same species, ``Sp.1/2'' refers to the negative effect of the number of adults of species 1 on juvenile survival of species 2, and so on.}
\label{fig:betaaest}
\end{figure}

Interaction strength estimates were not unbiased: all density-dependence parameters $\alpha$ and $\beta$ were on average overestimated (\Cref{fig:alphaest,fig:betaaest}). However, the accuracy of parameter estimation was sensitive to the choice of priors, and estimates obtained using exponential priors ($\theta \sim \text{Exp}(1)$, \ie set of priors 1) were overall close to the true values (minor bias). Among the three sets of priors considered, the set of priors 1 was the closest to the true values and was the only one presenting satisfactory coverage (\Cref{fig:alphacoverage,fig:betacoverage}).

\begin{figure}[H]
\includegraphics[width=0.95\linewidth]{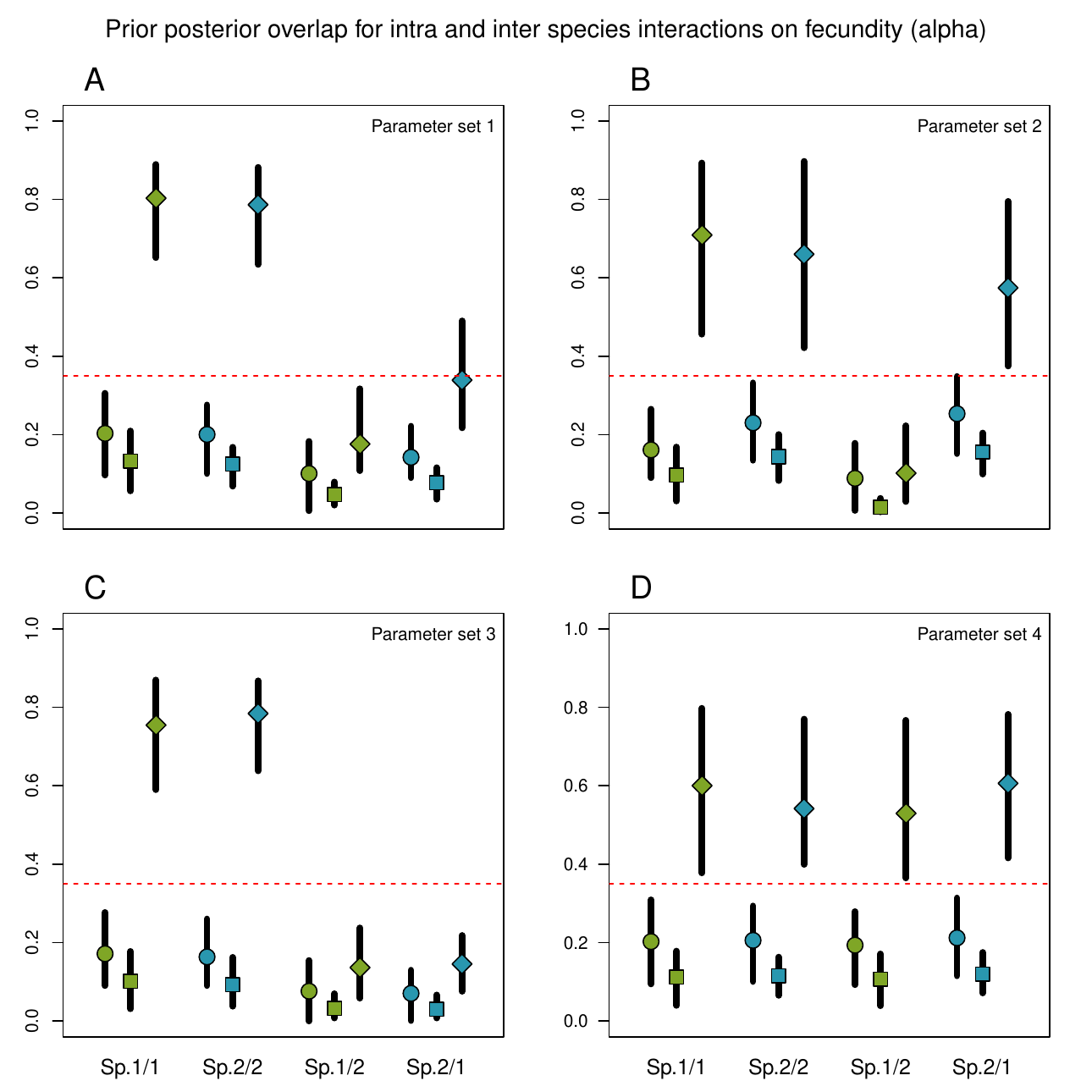}
\centering
\caption{Prior and posterior overlap for intra- and inter-species interactions on fecundity ($\alpha_{i,j}$) estimated under each parameter set (panel A = parameter set 1; panel B = parameter set 2; panel C = parameter set 3; panel D = parameter set 4). Vertical lines represent the 95\% intervals of the (typically 100) calculated prior and posterior overlaps, and open dots represent the mean (filled in green for effects of species 1 and in blue for effects of species 2). Round dots represent estimates when prior 1 was applied, square dots when prior 2 was applied, and diamond dots when prior 3 was applied to all $\alpha_{i,j}$ and $\beta_{i,j}$. ``Sp.1/1'' refers to $\alpha_{1,1}$, that is, the negative effect of the number of adults of species 1 on its own fecundity, ``Sp.1/2'' refers to the negative effect of the number of adults of species 1 on the fecundity of species 2, and so on. The red dotted lines represent the 35\% cutoff (values $>35\%$ suggesting weak identifiability of $\alpha_{i,j}$).}
\label{fig:alphaoverlap}
\end{figure}

\begin{figure}[H]
\includegraphics[width=0.95\linewidth]{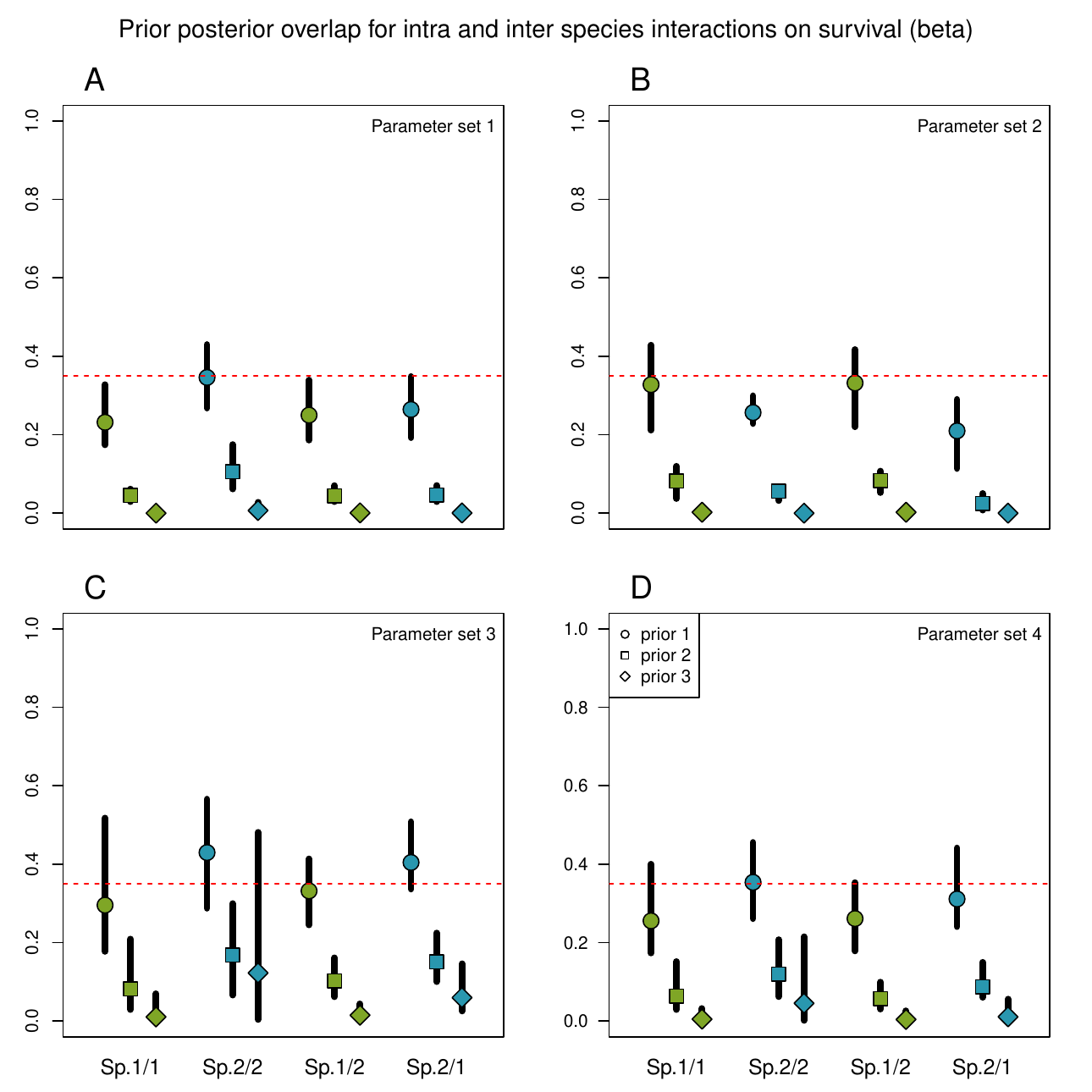}
\centering
\caption{Prior and posterior overlap for intra- and inter-species interactions on survival ($\beta_{i,j}$) estimated under each parameter set (panel A = parameter set 1; panel B = parameter set 2; panel C = parameter set 3; panel D = parameter set 4). Vertical lines represent the 95\% intervals of the (typically 100) calculated prior and posterior overlaps, and open dots represent the mean (filled in green for effects of species 1 and in blue for effects of species 2). Round dots represent estimates when prior 1 was applied, square dots when prior 2 was applied, and diamond dots when prior 3 was applied to all $\alpha_{i,j}$ and $\beta_{i,j}$. ``Sp.1/1'' refers to $\beta_{1,1}$, that is, the negative effect of the number of adults of species 1 on the survival of juveniles of the same species, ``Sp.1/2'' refers to the negative effect of the number of adults of species 1 on juvenile survival of species 2, and so on. The red dotted lines represent the 35\% cutoff (values $>35\%$ suggesting weak identifiability of $\beta_{i,j}$).}
\label{fig:betaoverlap}
\end{figure}

The sensitivity of these estimates to prior specification was also visible when investigating parameter identifiability via prior-posterior overlap diagnostics. More specifically, while prior-posterior overlaps for $\alpha$ parameters (density-dependence on fecundity) were relatively low when using prior 1 and prior 2, suggesting that these parameters were identifiable, this overlap was typically high when using the prior set 3 ($\alpha \sim \mathcal{LN}(\log(0.8)+0.05,0.05)$, suggesting weak identifiability (\Cref{fig:alphaoverlap}). Since this prior distribution mode is located far away from the true values of $\alpha$ (\Cref{fig:priors}), these overlaps indicate a strong influence of prior 3 on $\alpha$ estimates. For $\beta$ parameters however (\ie density-dependence on survival), the overlap was typically highest for prior 1, sometimes being higher than the classical 35\% cutoff \citep{garrett2000latent,gimenez2009weak} suggesting weak identifiability (\Cref{fig:betaoverlap}). These overlaps may be due to the low precision of parameter estimates, as true values fall well within the distribution of prior 1 (\Cref{fig:priors}).

\subsection{Estimating invasion criteria and competitive outcomes}
\begin{figure}[H]
\includegraphics[width=0.95\linewidth]{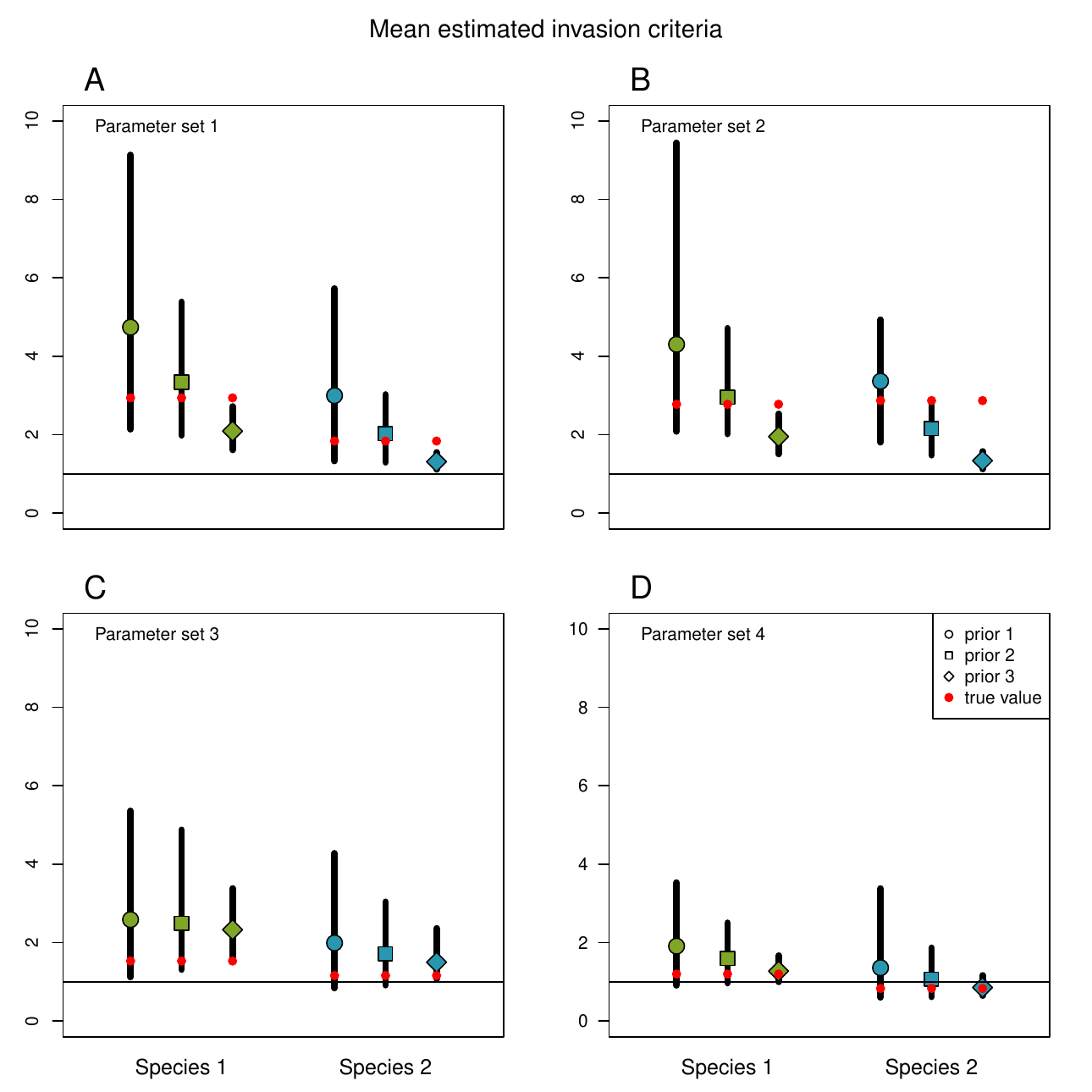}
\centering
\caption{Point estimates (\ie posterior means) of invasion criteria ($\mathcal{I}$) of species 1 (in green) and species 2 (in blue) estimated under each parameter set (panel A = parameter set 1; panel B = parameter set 2; panel C = parameter set 3; panel D = parameter set 4). Vertical lines represent the 95\% intervals of the (typically 100) posterior means, and open dots represent their means. Round dots represent estimates when prior 1 was applied, square dots when prior 2 was applied, and diamond dots when prior 3 was applied to all $\alpha_{i,j}$ and $\beta_{i,j}$. Red circles represent the true values.}
\label{fig:invasions}
\end{figure}

Unlike density-dependence parameters, invasion criteria were not systematically overestimated (\Cref{fig:invasions}), and the accuracy of parameter estimation was less sensitive to the choice of priors. More specifically, estimates obtained using ($\theta \sim \mathcal{L N}(0.5,1)$, \ie set of priors 2) were overall closest to true values and more precise, and both prior sets 1 and 2 presented satisfactory coverage (\Cref{fig:invasioncoverage}).

\begin{figure}[H]
\includegraphics[width=0.95\linewidth]{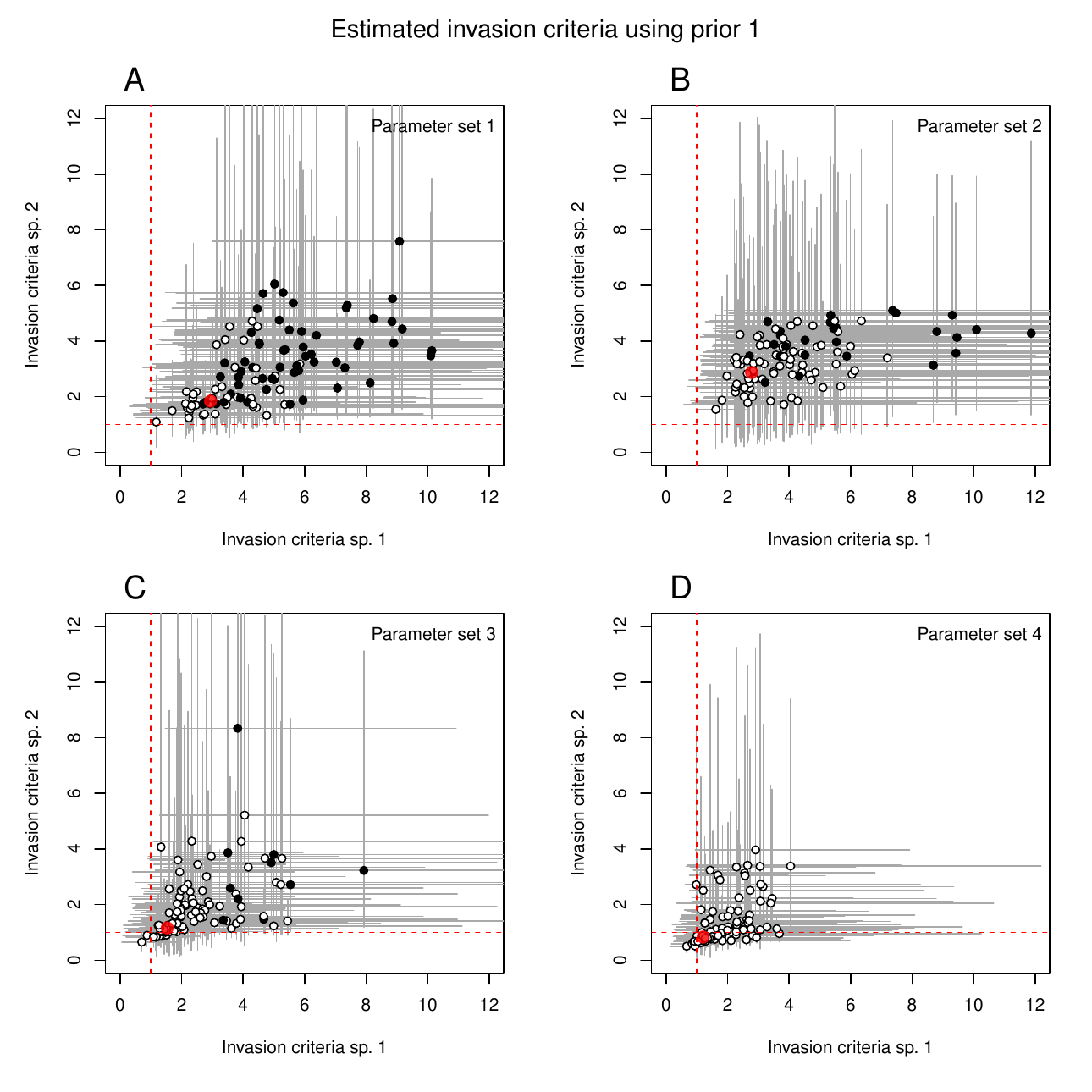}
\centering
\caption{Outcome of competition indicated by comparing the estimated invasion criterion of species 1 ($\mathcal{I}_1$) to those of species 2 ($\mathcal{I}_2$) for the four parameter sets when prior 1 was applied on interaction parameters. Dots represent the point estimates of the $\mathcal{I}_1$ $\mathcal{I}_2$ combinations and grey segments their 95 \% credible intervals. Red dotted lines represent the cut-off of 1 determining competition outcome: if both $\mathcal{I}_1>1$ and $\mathcal{I}_2>1$, then species are expected to coexist. If $\mathcal{I}_1>1$ and $\mathcal{I}_2<1$ then species 1 is expected to exclude species 2 (and \emph{vice versa}), and if both $\mathcal{I}_1<1$ and $\mathcal{I}_2<1$ a priority effect is expected. Red dots represent the true values. White dots represent ``uncertain'' outcomes, that is, $\mathcal{I}_1$ $\mathcal{I}_2$ pairs for which at least one of the two credible intervals (\ie for $\mathcal{I}_1$ and/or $\mathcal{I}_2)$ overlap 1. Black dots represent ``certain'' outcomes with no credible intervals spanning 1.}
\label{fig:inv2Dexp}
\end{figure}

Estimated competitive outcomes based on the invasion criteria of both species were often uncertain, but never incorrect when taking uncertainty into account (based on the 95\% credible intervals)(\Cref{fig:inv2Dexp,fig:inv2Dlognormlow,table:competoutcomes}). This was not true when ignoring uncertainty (\ie when only looking at point estimates) where the competitive outcome was wrong some times for parameter set 3, and about half of the time for parameter set 4, often estimating species co-existence when in fact species 1 won the competition (\Cref{fig:inv2Dexp,fig:inv2Dlognormlow,table:competoutcomes}).

\begin{table}[H]
\caption{Summary of the outcome of competition from all simulated datasets indicated by comparing the estimated invasion criterion of species 1 ($\mathcal{I}_1$) to those of species 2 ($\mathcal{I}_2$) for the four parameter sets and the three sets of priors, either when ignoring uncertainty (Uncertainty = No, \ie when only looking at point estimates) or when accounting for uncertainty (Uncertainty = Yes, \ie considering whether 95 \% credible intervals of $\mathcal{I}_1$ and $\mathcal{I}_2$ overlap with 1 or not). The number of simulations in \textbf{bold}  are those for which the outcome of competition is correct.
}\label{table:competoutcomes}
\centering
\begin{tabular}{llllllll}
  \hline
Prior & Parameter set & Uncertainty & Coexist & Sp.1 wins & Sp.2 wins & Priority & Uncertain \\ 
  \hline
1 & 1 & No & \bf{100} & 0 & 0 & 0 & 0 \\ 
  1 & 1 & Yes & \bf{57} & 0 & 0 & 0 & 43 \\ 
  1 & 2 & No & \bf{99} & 0 & 0 & 0 & 0 \\ 
  1 & 2 & Yes & \bf{26} & 0 & 0 & 0 & 73 \\ 
  1 & 3 & No & \bf{91} & 7 & 0 & 2 & 0 \\ 
  1 & 3 & Yes & \bf{10} & 0 & 0 & 0 & 90 \\ 
  1 & 4 & No & 50 & \bf{42} & 1 & 6 & 0 \\ 
  1 & 4 & Yes & 0 & \bf{0} & 0 & 0 & 99 \\
   \hline
  2 & 1 & No & \bf{99} & 0 & 0 & 0 & 0 \\ 
  2 & 1 & Yes & \bf{55} & 0 & 0 & 0 & 44 \\ 
  2 & 2 & No & \bf{100} & 0 & 0 & 0 & 0 \\ 
  2 & 2 & Yes & \bf{22} & 0 & 0 & 0 & 78 \\ 
  2 & 3 & No & \bf{93} & 7 & 0 & 0 & 0 \\ 
  2 & 3 & Yes & \bf{14} & 0 & 0 & 0 & 86 \\ 
  2 & 4 & No & 46 & \bf{48} & 1 & 4 & 0 \\ 
  2 & 4 & Yes & 0 & \bf{1} & 0 & 0 & 98 \\
   \hline
  3 & 1 & No & \bf{100} & 0 & 0 & 0 & 0 \\ 
  3 & 1 & Yes & \bf{28} & 0 & 0 & 0 & 72 \\ 
  3 & 2 & No & \bf{99} & 0 & 0 & 0 & 0 \\ 
  3 & 2 & Yes & \bf{23} & 0 & 0 & 0 & 76 \\ 
  3 & 3 & No & \bf{99} & 1 & 0 & 0 & 0 \\ 
  3 & 3 & Yes & \bf{51} & 0 & 0 & 0 & 49 \\ 
  3 & 4 & No & 14 & \bf{84} & 2 & 0 & 0 \\ 
  3 & 4 & Yes & 0 & \bf{6} & 0 & 0 & 94 \\ 
   \hline
\end{tabular}
\end{table}

\section{Discussion}

We fitted a stage-structured competition IPM to three data sources (counts, capture-mark-recapture histories, fecundities), using density-dependent vital rates to model interactions between species. Interaction coefficients could be correctly estimated, albeit with a small bias, but this required using a fairly uninformative prior for interaction coefficients (the maximum entropy prior). 

Invasion criteria, which could be derived here due to the amenability of this model to analytical calculations, were less sensitive to prior choice than interaction parameters, both in terms of bias and coverage. This echoes results of \citet{kao2018practical} who found that transmission parameters, the equivalents of our density-dependent parameters, were not practically identifiable in an epidemiological model, but $R_0$, analogous to our invasion criteria, was identifiable. We also attempted to go one step deeper and tried to identify the pathways leading to coexistence or exclusion. More specifically, we tried to distinguish the role of competition on fecundity and competition on survival in producing the observed competitive outcomes, by calculating additional invasion criteria (\citealt{bardon2023effects}, see Supplementary Section \ref{SuppB}). However, these derived invasion parameters appeared particularly hard to estimate (\Cref{fig:Rinvasions}).

That said, accounting for uncertainty in invasion criteria never led to wrongly estimated competitive outcomes, while posterior means (point estimates) did predict incorrect competitive outcomes a few times (\ie 1--7 \% of simulations depending on the prior choice) for parameter set 3, for which species coexist, and often (\ie 16--58 \%) for parameter set 4 for which one species wins the competition (\Cref{table:competoutcomes}). These findings tend to confirm suggestions by \citet{armitage2024remain,bowler2022accounting} that it is critical to place uncertainty estimates around point estimates of coexistence criteria. 

As explained above, we had more difficulty pinpointing extinction trajectories generated by one species winning the competition over the other (parameter set 4). In this case, it was more difficult to find competitive outcome criteria intervals within the extinction regions (\Cref{fig:inv2Dexp}). If this bias is also present in empirical articles based on real data, our results suggest that some bias toward prediction of coexisting species may be present, or might occur in the future in the literature. This is interesting as plant ecology field studies combined with community-level statistical modelling have typically a higher tendency to see coexistence than do experimental designs \citep{adler2018weak}. The difficulty in predicting the extinction of one population and a single winner could be due to the fact that for most of the time span studied, both species actually coexist even though one of them is \textit{headed} towards extinction. A framework such as ours, which accounts for uncertainty, has at least the advantage to flag these extinction scenarios as ``uncertain'' rather than ``coexistence'', as would occur when neglecting parameter uncertainties. Further work might focus on modelling coupled populations where one of them actually reaches zero individuals and stays there for a long-time, which requires some modifications to the current framework. We suspect that it could be particularly interesting when done in a spatial setting, with extinction occurring at different times in different places, which could increase the quantity of information available to the model. One should also note that invasion growth rates, which predict a deterministically-driven path to extinction whenever below unity, may not account for all pathways to extinction, which always occurs eventually in models with demographic stochasticity, with chance extinctions getting more and more likely as population size drops. That said, invasion growth rates have been typically found useful in predicting coexistence or extinction even in systems with demographic stochasticity \citep{schreiber2023does}. 

The framework chosen here, with theory-driven functional forms (Beverton-Holt), allows to compute analytical invasion criteria which are useful to predict coexistence or other competitive outcomes. Other ways to specify IPMs use classically log and logit links \citep{peron2012integrated,queroue2021multispecies} but these do not allow the computation of invasion criteria, as the algebra becomes too complicated. Using Ricker rather than Beverton-Holt functions, which is equivalent to a log link in functions relating vital rates to densities, it may not be possible to obtain analytical invasion criteria except in some special cases. 
However, it is possible that models with log and logit links behave better in other ways and that interaction coefficients could be better estimated in these models. For instance, we encountered identifiability issues with fertility and survival max parameters $\pi$ and $\phi$, this might or might not occur in models with different choices of functional forms \citep{barraquand2019integrating,paquet2022assessing}. As the models with log- and logit links cannot yield directly mathematical indicators of competitive outcomes, one might in the future compare the precision of predicted community trajectories from those statistically-driven models vs the theory-driven models constructed here. Yet another strategy is to construct two time-dependent (but density-independent) one-species IPMs, deduce from these the variation over time of growth rates and population sizes of both species, and then construct more phenomenological Lotka--Volterra-style models, as done by \citet{gamelon2019accounting}. How the latter post-hoc regression strategy, which appears handy but a bit less conventional---from a statistical standpoint, compared to directly estimating the density-dependencies of vital rates---relates to theoretical coexistence criteria for stage-structured systems is currently unknown. 

The quantity of data may influence, of course, our ability to recover interaction parameters. However, our 30 year survey with counts each year and 100 marked juveniles each year, as well as 50 broods/litters surveyed per year could certainly be considered a large sample size for animal populations. Exceptionally a single-species dataset might go beyond that \citep[e.g. in well-studied and abundant birds][]{clark2023climate} but then to fit a multispecies model we need multiple species, and having all species with more data than considered here might be unusual. Spatially replicated surveys could hold some promise to increase sample size. One difficulty is then the interactions themselves, like all parameters, can vary spatially but some hierarchical modelling could be attempted. Plant surveys might hold the most promise to attain very large sample sizes through spatial replication. Although the capture-recapture models used here for survival modelling may look a bit foreign for plants to some readers, we note that several authors have argued that they could be actually very useful to properly account for imperfect detection of individuals \citep{kery2003effects,martinez2024estimation}. 

In this paper, we have shown how to systematically propagate uncertainty in the prediction of competitive outcomes in stage-structured systems, be those coexistence, exclusion or priority effects, in a Bayesian framework. This was done by combining multiple data sources in a two-species Integrated Population Model, whose deterministic skeleton is sufficiently analytically tractable to derive invasion criteria. The model framework was very successful in pinpointing coexistence scenarios. Exclusion was more difficult to ascertain, though our framework allows at least to tell what we can or cannot predict. Future work might focus on these competitive exclusion cases, and ways to estimate by which mechanistic pathways coexistence can be brought about. 

\section*{Acknowledgements}

Colleagues from IMB and Biogeco labs are thanked for a welcoming environment.

\section*{Data and code}

Computer R code used to generate and analyze the data is available at \url{https://github.com/MatthieuPaquet/IPM_competition_2_species_age_structured}

\section*{Funding}

Funding was provided through grant ANR-20-CE45-0004 to FB.

\bibliographystyle{ecol_let}
\bibliography{mybiblio}
\pagebreak
\renewcommand\thesection{S}
\setcounter{figure}{0}
\renewcommand{\thefigure}{S\arabic{figure}}
\setcounter{table}{0}
\renewcommand{\thetable}{S\arabic{table}}
\section*{Supplementary Information}
\renewcommand\thesubsection{\Alph{subsection}}
\subsection{Supplementary figures} \label{SuppA}
\begin{figure}[H]
\includegraphics[width=0.95\linewidth]{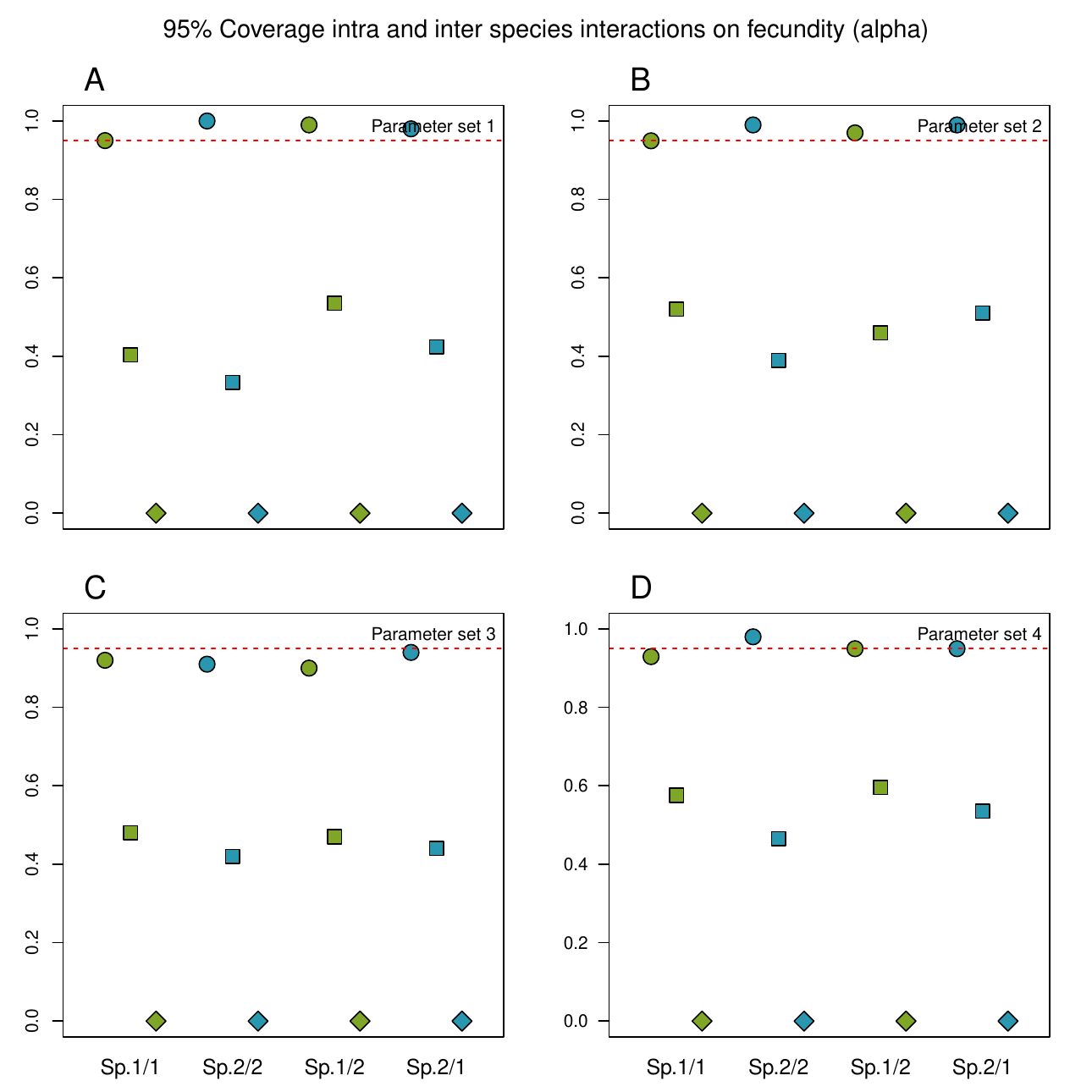}
\centering
\caption{Coverages (= proportion of simulations where 95\% CrI of estimated parameter include the true parameter value) for intra- and inter-species interactions on fecundity ($\alpha_{i,j}$) estimated under each parameter set (panel A = parameter set 1; panel B = parameter set 2; panel C = parameter set 3; panel D = parameter set 4). Green dots represent effects of species 1 and blue dots represent effects of species 2). Round dots represent coverages when prior 1 was applied, square dots when prior 2 was applied, and diamond dots when prior 3 was applied, to all $\alpha_{i,j}$ and $\beta_{i,j}$. ``Sp.1/1'' refers to $\alpha_{1,1}$, that is, the negative effect of the number of adults of species 1 on its own fecundity, ``Sp.1/2'' refers to the negative effect of the number of adults of species 1 on the fecundity of species 2, and so on. The red dotted lines indicate a coverage of 95\%.}
\label{fig:alphacoverage}
\end{figure}

\begin{figure}[H]
\includegraphics[width=0.95\linewidth]{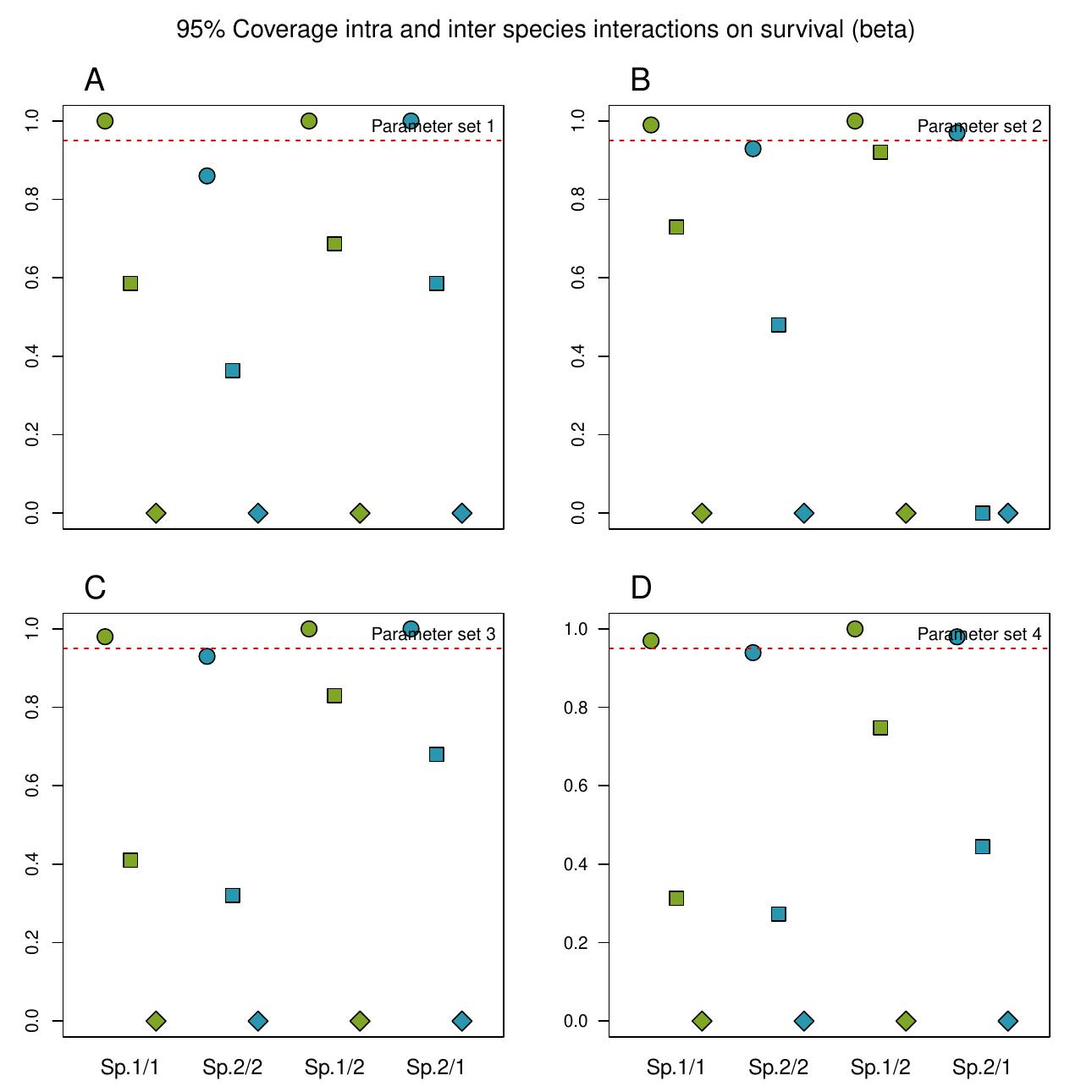}
\centering
\caption{Coverages (= proportion of simulations where 95\% CrI of estimated parameter include the true parameter value) for intra- and inter-species interactions on survival ($\beta_{i,j}$) estimated under each parameter set (panel A = parameter set 1; panel B = parameter set 2; panel C = parameter set 3; panel D = parameter set 4). Green dots represent effects of species 1 and blue dots represent effects of species 2). Round dots represent coverages when prior 1 was applied, square dots when prior 2 was applied, and diamond dots when prior 3 was applied, to all $\alpha_{i,j}$ and $\beta_{i,j}$. ``Sp.1/1'' refers to $\beta_{1,1}$, that is, the negative effect of the number of adults of species 1 on the survival of juveniles of the same species, ``Sp.1/2'' refers to the negative effect of the number of adults of species 1 on juvenile survival of species 2, and so on. The red dotted lines indicate a coverage of 95\%}
\label{fig:betacoverage}
\end{figure}

\begin{figure}[H]
\includegraphics[width=0.95\linewidth]{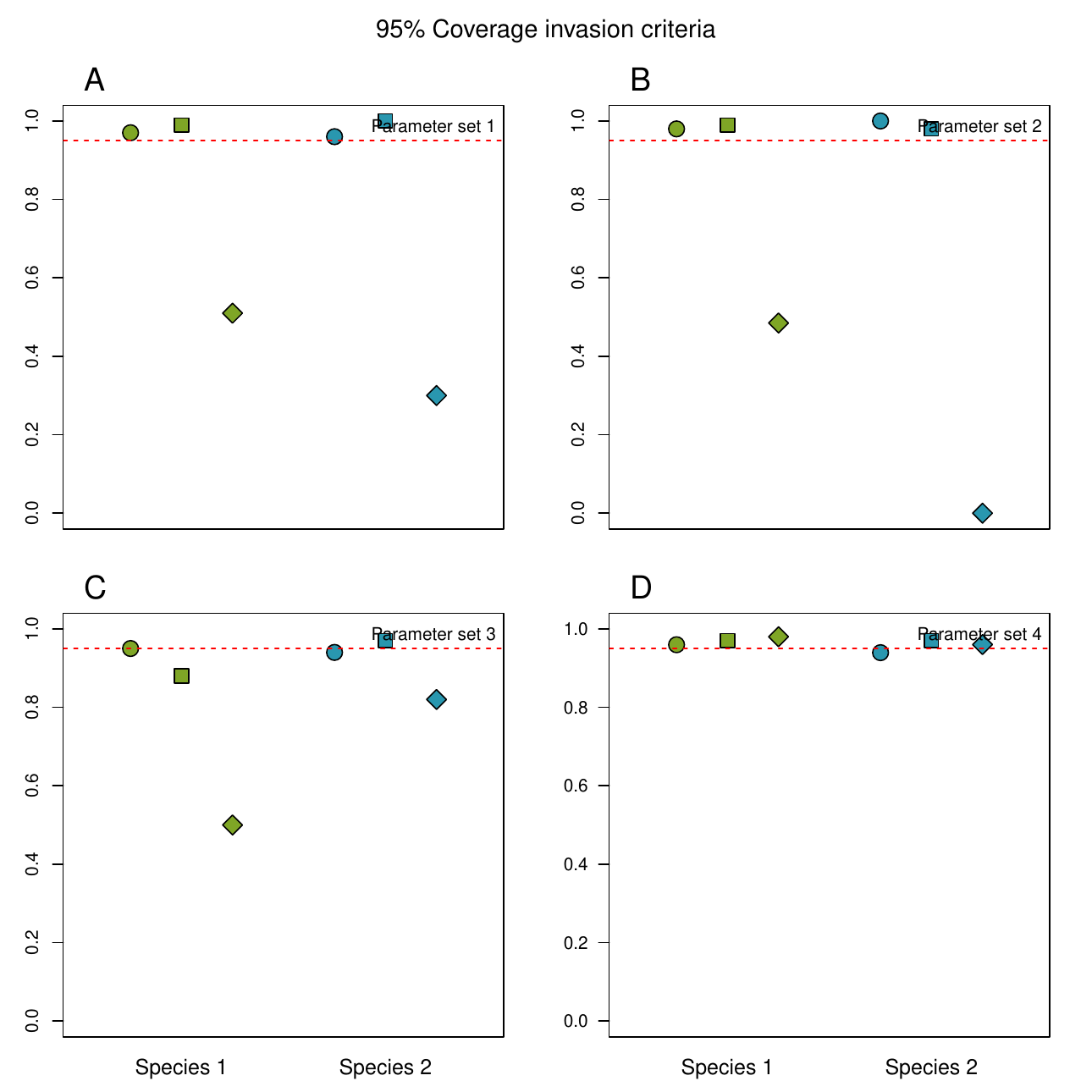}
\centering
\caption{Coverages (= proportion of simulations where 95\% CrI of estimated parameter include the true parameter value) for invasion criteria ($\mathcal{I}$) estimated under each parameter set (panel A = parameter set 1; panel B = parameter set 2; panel C = parameter set 3; panel D = parameter set 4). Green dots represent $\mathcal{I}$ of species 1 and blue dots represent $\mathcal{I}$ of species 2). Round dots represent coverages when prior 1 was applied, square dots when prior 2 was applied, and diamond dots when prior 3 was applied, to all $\alpha_{i,j}$ and $\beta_{i,j}$. The red dotted lines indicate a coverage of 95\%.}
\label{fig:invasioncoverage}
\end{figure}
\begin{figure}[H]
\includegraphics[width=0.95\linewidth]{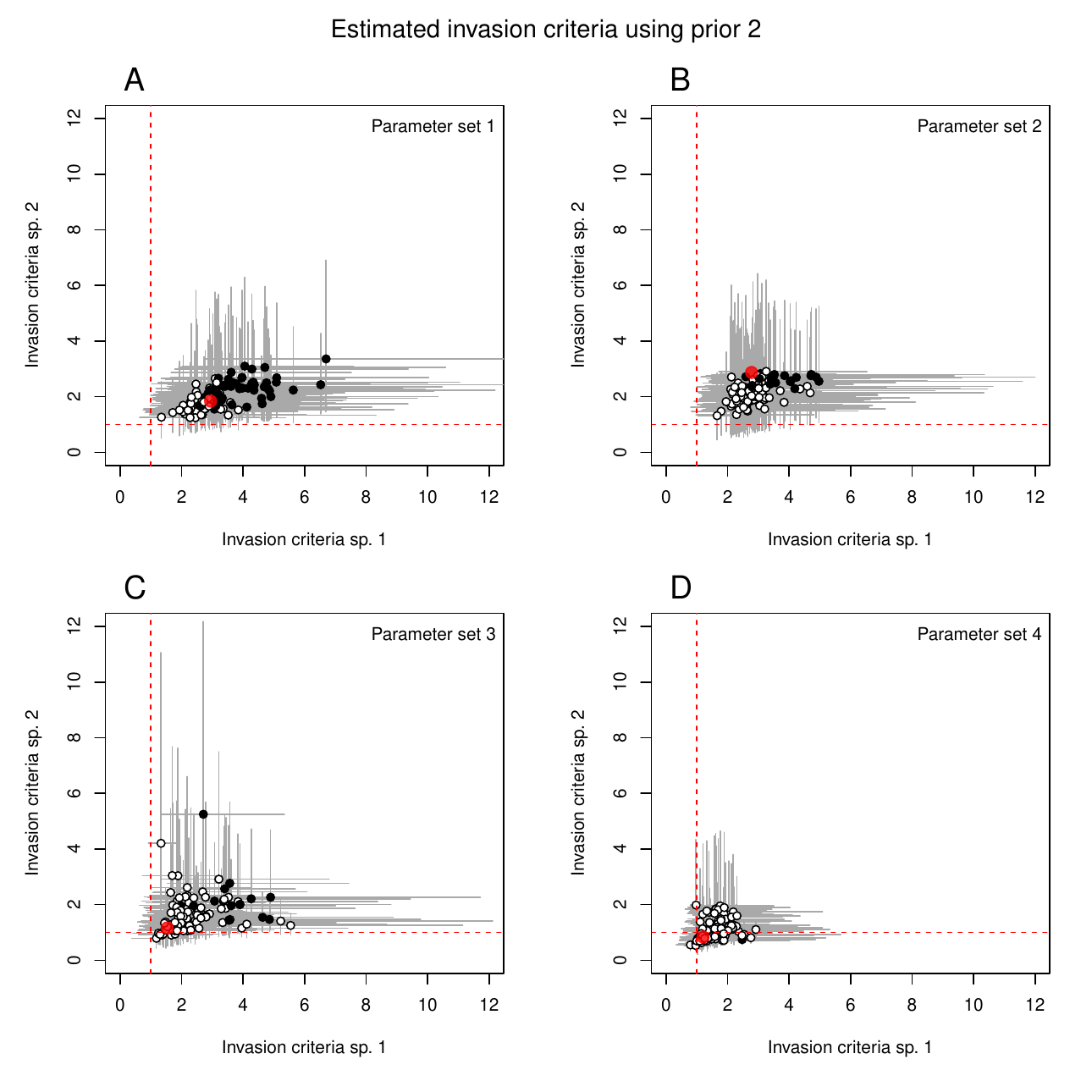}
\centering
\caption{Outcome of competition indicated by comparing the estimated invasion criterion of species 1 ($\mathcal{I}_1$) to those of species 2 ($\mathcal{I}_2$) for the four parameter sets when prior 2 was applied on interaction parameters. Dots represent the point estimates of the $\mathcal{I}_1$ $\mathcal{I}_2$ combinations and grey segments their 95 \% credible intervals. Red dotted lines represent the cut-off of 1 determining competition outcome: if both $\mathcal{I}_1>1$ and $\mathcal{I}_2>1$, then species are expected to coexist. If $\mathcal{I}_1>1$ and $\mathcal{I}_2<1$ then species 1 is expected to exclude species 2 (and \emph{vice versa}), and if both $\mathcal{I}_1<1$ and $\mathcal{I}_2<1$ a priority effect is expected. Red dots represent the true values. White dots represent ``uncertain'' outcomes, that is, $\mathcal{I}_1$ $\mathcal{I}_2$ pairs for which at least one of the two credible intervals (\ie for $\mathcal{I}_1$ and/or $\mathcal{I}_2)$ overlap 1. Black dots represent ``certain'' outcomes with no credible intervals spanning 1.}
\label{fig:inv2Dlognormlow}
\end{figure}

\subsection{Computation of invasion criteria from the fecundity and survival submodels} \label{SuppB}

To try and reveal the coexistence pathways (whether it is due to competition on fecundity or on survival) we computed the invasion criteria given by \citet{fujiwara2011coexistence} as used in \citet{bardon2023effects}. For the fecundity submodel, the invasion criteria for species 1 is
given by:
\begin{equation}
\mathcal{R}_{\alpha 1} = \frac{R'_{\alpha1}}{R'_{\alpha2}} \frac{\alpha_{22}}{\alpha_{12}} 
\end{equation}
with
\begin{equation}
  R'_{\alpha1}=\frac{\pi_{1} }{ f_{1}^{(c)} } - 1
\end{equation}
and
\begin{equation}
  f_1^{(c)} = \frac{1}{\gamma s_{1j}}(1 - s_{1a}) \left(1- s_{1j}+ \gamma s_{1j}\right) = \frac{1}{s_{1j}}(1 - s_{1a}) 
\end{equation}
since $\gamma=1$ in our study.

Similarly, for the survival submodel, the invasion criteria for species 1 is:

\begin{equation}
\mathcal{R}_{\beta1} = \frac{R'_{\beta1}}{R'_{\beta2}} \frac{\beta_{22}}{\beta_{12}} 
\end{equation}
with
\begin{equation}
    R'_{\beta1}=\frac{\phi_{1} }{ s_{1}^{(c)} }  - 1
\end{equation}
and
\begin{equation}
      s_{1}^{(c)} = \frac{1 - s_{1a}}{(1-\gamma)(1 - s_{1a}) + \pi_1\gamma}= \frac{1 - s_{1a}}{\pi_1}
\end{equation}

\begin{table}[H]
\caption{Values of invasion criteria of the fecundity and survival submodels for each parameter set
}\label{table:Rinvasions}
\centering
\begin{tabular}{llllll}
  \hline
Parameter set & $\mathcal{R}_{\alpha1}$ & $\mathcal{R}_{\alpha2}$ & $\mathcal{R}_{\beta1}$ & $\mathcal{R}_{\beta2}$ & Outcome \\ 
  \hline
1 & \bf{2.42} & \bf{1.38} & \bf{2.01} & \bf{1.38} & Coexistence\\ 
2 & \bf{6.04} & 0.74 & 0.97 & \bf{8.28} & Emergent coexistence\\ 
3 & \bf{2.81} & \bf{2.36} & 0.78 & 0.50 & Emergent coexistence\\ 
4 & \bf{1.21} & 0.83 & \bf{1.21} & 0.83 & Sp.1 wins\\ 
   \hline
\end{tabular}
\end{table}
\begin{figure}[H]
\includegraphics[width=0.95\linewidth]{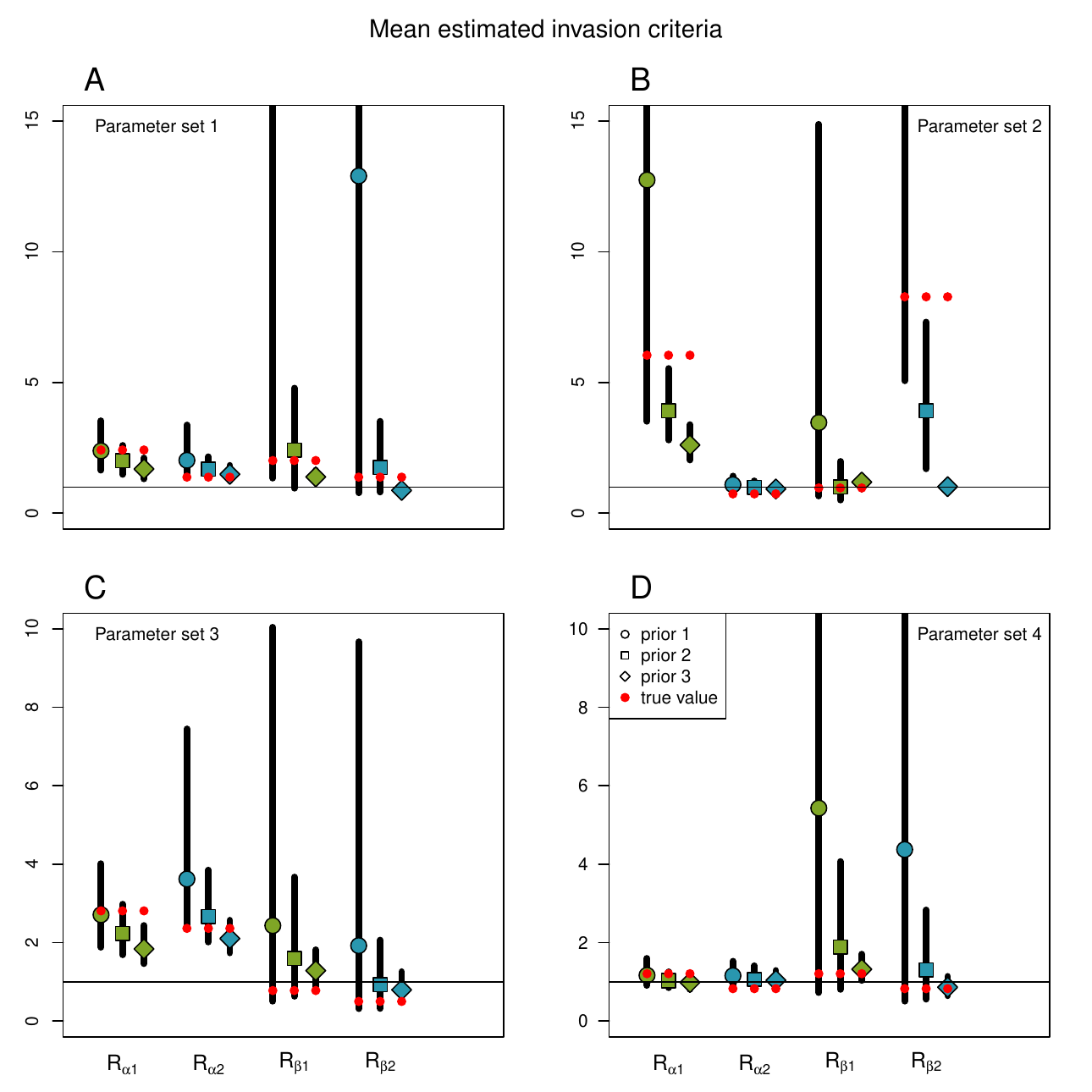}
\centering
\caption{Point estimates (\ie posterior means) of invasion criteria for fecundity ($\mathcal{R}_{\alpha}$) and for survival ($\mathcal{R}_{\beta}$) of species 1 (in green) and species 2 (in blue) estimated under each parameter set (panel A = parameter set 1; panel B = parameter set 2; panel C = parameter set 3; panel D = parameter set 4). Vertical lines represent the 95\% intervals of the (typically 100) posterior means, and open dots represent their means. Round dots represent estimates when prior 1 was applied, square dots when prior 2 was applied, and diamond dots when prior 3 was applied, to all $\alpha_{i,j}$ and $\beta_{i,j}$. Red circles represent the true values.}
\label{fig:Rinvasions}
\end{figure}

\begin{figure}[H]
\includegraphics[width=0.95\linewidth]{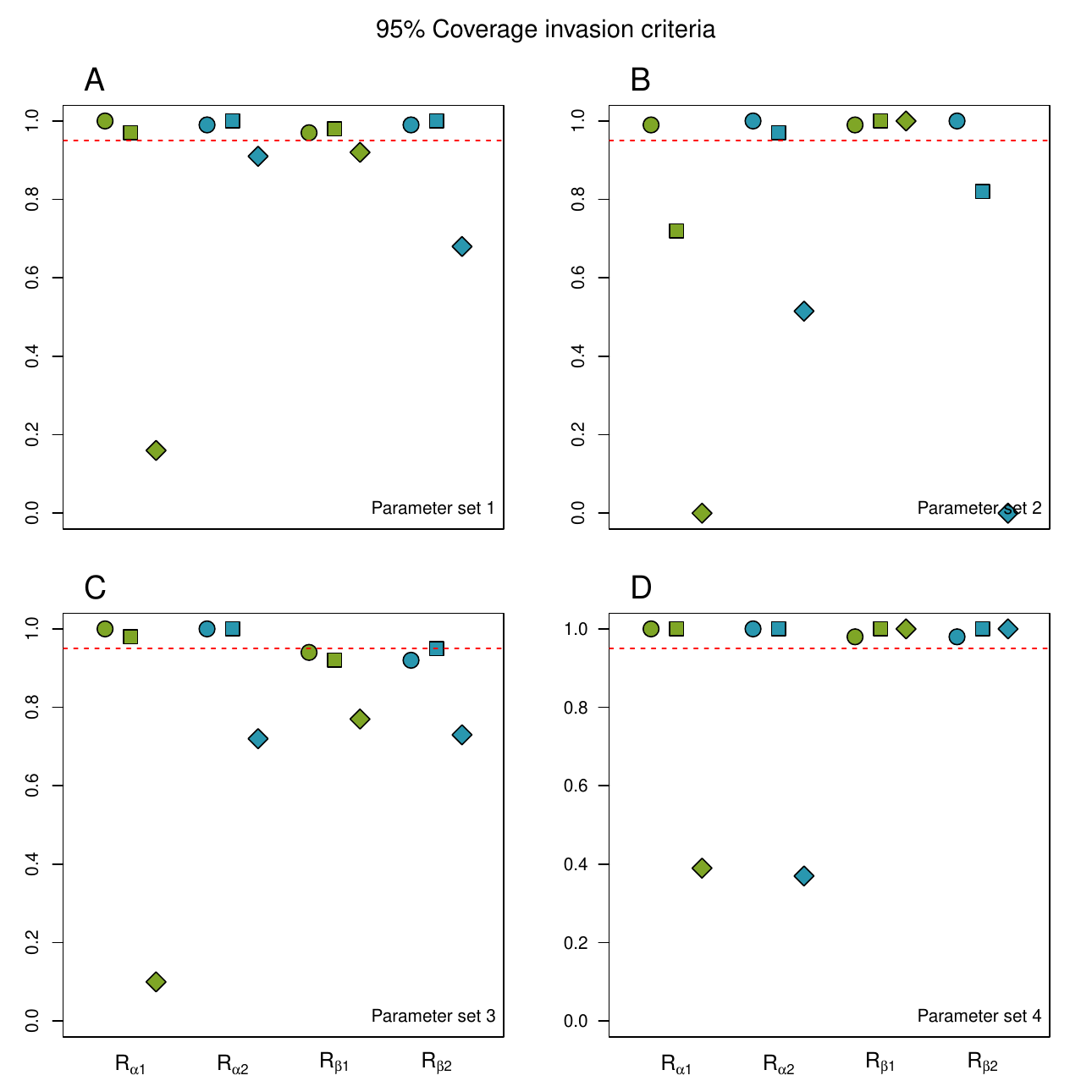}
\centering
\caption{Coverages (= proportion of simulations where 95\% CrI of estimated parameter includes the true parameter value) for the invasion criteria of the fecundity and survival submodels ($\mathcal{R}_{\alpha}$ and $\mathcal{R}_{\beta}$) estimated under each parameter set (panel A = parameter set 1; panel B = parameter set 2; panel C = parameter set 3; panel D = parameter set 4). Green dots represent$\mathcal{R}$ of species 1 and blue dots represent $\mathcal{R}$ of species 2). Round dots represent coverages when prior 1 was applied, square dots when prior 2 was applied, and diamond dots when prior 3 was applied, to all $\alpha_{i,j}$ and $\beta_{i,j}$. The red dotted lines indicate a coverage of 95\%.}
\label{fig:Rinvasioncoverage}
\end{figure}

\end{document}